\documentclass[aps,superscriptaddress,twoside,twocolumn,floatfix,a4paper,longbibliography]{revtex4-1}

\usepackage{graphicx}
\usepackage{amsmath}
\usepackage{dcolumn}   % needed for some tables
\usepackage{bm}        % for math
\usepackage{amssymb}   % for math
\usepackage{graphicx,epsfig}
\usepackage{color}
\usepackage[usenames,dvipsnames]{xcolor}
\usepackage[ocgcolorlinks,colorlinks=true,linkcolor=blue,citecolor=red]{hyperref}
\usepackage{amsmath,amssymb, amsthm, amsfonts}
\usepackage{xspace}
\usepackage{xcolor}
\usepackage{verbatim}
\usepackage{mathtools}

\begin{document}
\title{Hypersensitive tunable Josephson escape sensor for gigahertz astronomy}

%Authors
\author{Federico Paolucci}
\email{federico.paolucci@nano.cnr.it}
\affiliation{INFN Sezione di Pisa, Largo Bruno Pontecorvo, 3, I-56127 Pisa, Italy}
\affiliation{NEST, Istituto Nanoscienze-CNR and Scuola Normale Superiore, I-56127 Pisa, Italy}

\author{Nadia Ligato}
\affiliation{NEST, Istituto Nanoscienze-CNR and Scuola Normale Superiore, I-56127 Pisa, Italy}

\author{Vittorio Buccheri}
\affiliation{INFN Sezione di Pisa, Largo Bruno Pontecorvo, 3, I-56127 Pisa, Italy}
\affiliation{NEST, Istituto Nanoscienze-CNR and Scuola Normale Superiore, I-56127 Pisa, Italy}

\author{Gaia Germanese} 
\affiliation{NEST, Istituto Nanoscienze-CNR and Scuola Normale Superiore, I-56127 Pisa, Italy}
\affiliation{Dipartimento di Fisica dell'Universit\`a di Pisa, Largo Pontecorvo 3, I-56127 Pisa, Italy}

\author{Pauli Virtanen} 
\affiliation{University of Jyvaskyla, Department of Physics and Nanoscience Center, P.O. Box 35 (YFL), FI-40014 University of Jyv\"askyl\"a, Finland}

\author{Francesco Giazotto}
\email{francesco.giazotto@sns.it}
\affiliation{NEST, Istituto Nanoscienze-CNR and Scuola Normale Superiore, I-56127 Pisa, Italy}

%%%%%%%%%%%%%%%%%%%%%%%%%%%%%%%%%%%%%%%%%%%%%%%%%%%%%%%%%%%%%%%%%%%%%
%% The document title should be given as usual
%% A short title can be given as a *suggestion* for running headers.
%%%%%%%%%%%%%%%%%%%%%%%%%%%%%%%%%%%%%%%%%%%%%%%%%%%%%%%%%%%%%%%%%%%%%

%%%%%%%%%%%%%%%%%%%%%%%%%%%%%%%%%%%%%%%%%%%%%%%%%%%%%%%%%%%%%%%%%%%%%
%% Start the main part of the manuscript here.
%%%%%%%%%%%%%%%%%%%%%%%%%%%%%%%%%%%%%%%%%%%%%%%%%%%%%%%%%%%%%%%%%%%%%

\begin{abstract}
 Single-photon detectors and bolometers represent the bridge between different topics in science, such as quantum computation, astronomy, particle physics and biology. Nowadays, superconducting bolometers and calorimeters are the most sensitive detectors in the THz and sub-THz bands.
 Here, we propose and demonstrate a Josephson escape sensor (JES) that could find natural application in astrophysics. The JES is composed of a fully superconducting one-dimensional Josephson junction, whose resistance versus temperature characteristics can be precisely controlled by a bias current. Therefore, differently from the traditional superconducting detectors, the JES sensitivity and working temperature can be in situ simply and finely tuned depending on the application requirements. Interestingly, a JES bolometer is expected to show an intrinsic thermal fluctuation noise equivalent power ($NEP$) of the order of $10^{-25}$ W/Hz$^{1/2}$, while a JES calorimeter could provide a frequency resolution of about 2 GHz, 
 as deduced from the experimental data. In addition, the sensor can operate at the critical temperature, i.e.,
 working as a conventional transition edge sensor (TES), with a $NEP\sim 6 \times 10^{-20}$ W/Hz$^{1/2}$ and a frequency resolution $\sim100$ GHz.   
\end{abstract}
\maketitle
\section{Introduction}

Sensitive photon detection in the gigahertz band constitutes the cornerstone to study different phenomena in astronomy \cite{Rowan2009}, such as radio burst sources \cite{Marcote2020}, galaxy formation \cite{Tabatabaei2017}, cosmic microwave background \cite{Sironi1999}, axions \cite{Capparelli2016,Irastorza2018}, comets \cite{Falchi1988}, gigahertz-peaked spectrum radio sources \cite{Odea1998} and supermassive black holes \cite{Issaoun2019}. 
Nowadays, state of the art detectors for astrophysics are mainly based on transition edge sensors \cite{Irwin1995,Irwin2006,Khosropanah2010} and kinetic inductance detectors \cite{Day2003,Visser2014,Monfardini2016}. 
Overall, most sensible nanobolometers so far are superconducting detectors \cite{Morgan2018} showing a noise-equivalent power (NEP) as low as $\sim 2\times 10^{-20}$ W/Hz$^{1/2}$ \cite{Kokkoniemi2019}. 
Yet, fast thermometry at the nanoscale was demonstrated as well with Josephson junctions through switching current measurements \cite{Zgirski2018,Wang2018}.
In general, detection performance are set by the fabrication process and limited by used materials.

In this work, we conceive and demonstrate a tunable Josephson escape sensor (JES) based on the precise current control of the temperature dependence of a fully superconducting one-dimensional nanowire  Josephson junction. 
The JES might be at the core of future hypersensitive \textit{in situ}-tunable bolometers or single-photon detectors working in the gigahertz regime.
Operated as a bolometer the JES points to a thermal fluctuation noise (TFN) NEP$_{TFN}\sim 1\times 10^{-25}$ W/Hz$^{1/2}$, which as a calorimeter bounds the frequency resolution above $\sim 2$ GHz, and resolving power below $\sim 40$ at $50$ GHz, as deduced from the experimental data.

Beyond the obvious applications in advanced ground-based \cite{Ade2019} and space \cite{Adams2020} telescopes for gigahertz astronomy, the JES might represent a breakthrough in several fields of quantum technologies ranging from  sub-THz communications \cite{Federici2010} and quantum computing \cite{Obrien2007} to cryptography \cite{Gisin2002} and quantum key distribution \cite{Tittel2019}.

\section{GHz astrophysics and JES operation principle}

\begin{figure} [t!]
        \begin{center}
                \includegraphics [width=0.9\columnwidth]{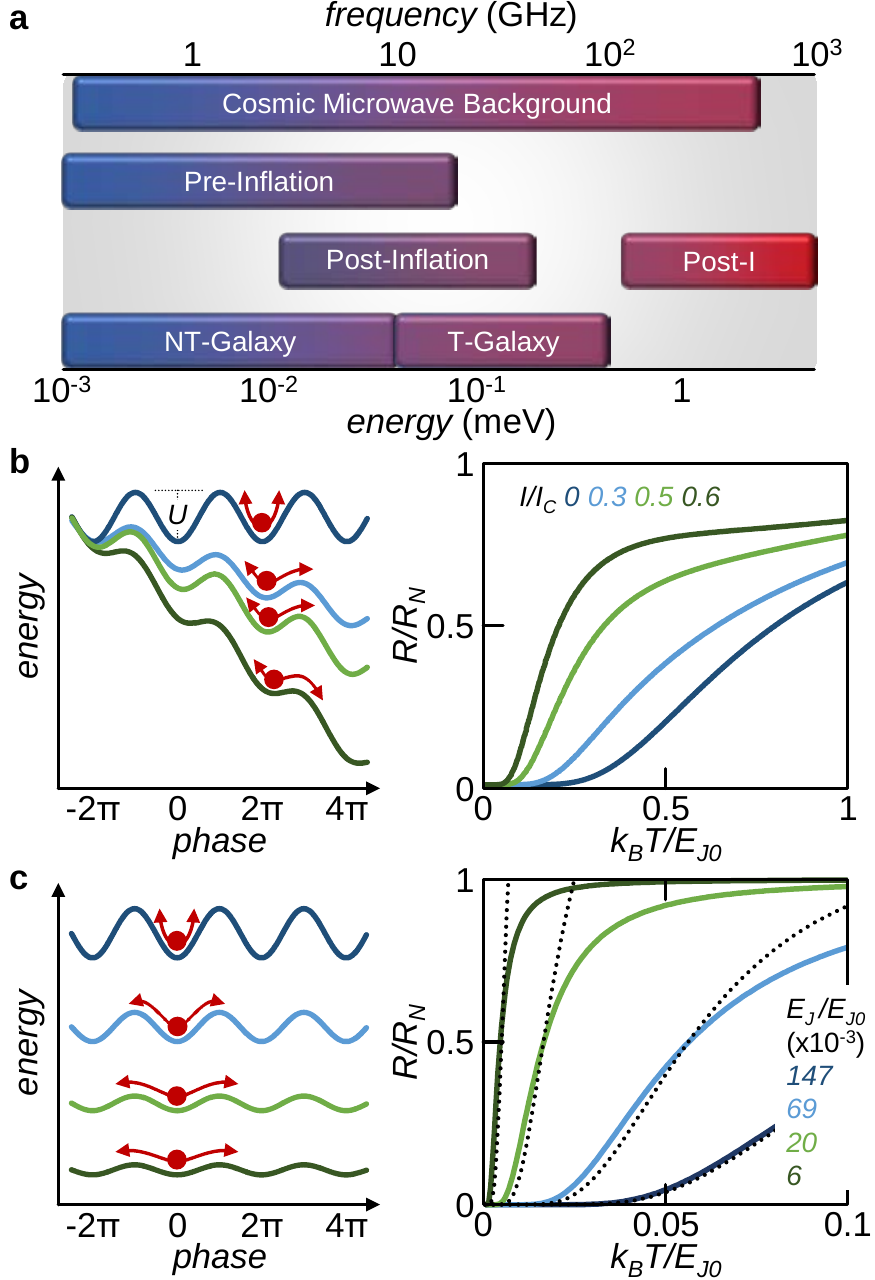}
        \end{center}
        \caption{\label{fig:Fig1}\textbf{GHz astronomy and JES operation principle.} \textbf{(a)} Frequency and energy ranges relevant for different astronomical investigations, such as cosmic microwave background \cite{Sironi1999}, pre-inflation and post-inflation axions \cite{Irastorza2018}, non-thermal (NT) and thermal (T) galaxy emission \cite{Tabatabaei2017}. 
\textbf{(b)} Left: Schematics of the washboard model versus phase difference across a JJ for different values of bias current 
($I$ rises from red to green with the same color code of the right panel).
$U$ represents the energy barrier for the escape of the phase particle 
(black dots).
Right: RSJ resistance ($R$) versus temperature ($T$) characteristics of a JJ for different values of $I$. 
Both the transition temperature and its width decrease by rising $I$. 
\textbf{(c)} Left: Schematics of the washboard model versus phase difference across a JJ for different values of the Josephson energy 
$E_J$ decreases from red to green with the same color code of the right panel).
Right: $R$ versus $T$ characteristics  of a JJ for different values of $E_J$ at $I=0$. 
Black dotted curves are calculated for $E_J=E_{J0}$ by varying $I$ in order to have the same transition temperature.}
\end{figure}

Many features of the universe are hidden in infrared and microwave faint signals \cite{Rowan2009}. 
In particular, the study of cosmic microwave background polarization \cite{Sironi1999,Seljak1997} and galaxy expansion \cite{Tabatabaei2017} benefits from ultrasensitive gigahertz bolometers, while the existence of axion-like-particles \cite{Irastorza2018,Capparelli2016} might be proven through the revelation of microwave single-photons (see Fig. \ref{fig:Fig1}-a). 

To improve photon detection sensitivity, novel superconducting sensors have been developed by \emph{miniaturizing} the active region \cite{Karasik2007,Wei2008}, and drastically \emph{lowering} their operation temperature via the Josephson coupling in complex nanostructures \cite{Kokkoniemi2019,Chen2011,Nararajan2012,Karimi2020,Virtanen2018,Giazotto2008, Kuzmin2019,Govenius2016}. 
Their properties are thus defined during the fabrication process and can not be tuned during the operation.
In analogy to the widespread transition edge sensor (TES), the JES exploits the change of resistance of a superconductor when transitioning to the dissipative state: either the absorption of radiation or the change of temperature \cite{Zgirski2018,Wang2018} trigger the passage from the superconducting to the normal regime yielding a sizable signal from the sensor.
In contrast to TESs, the JES benefits from the possibility to finely tune \textit{in situ} its working temperature and sensitivity.

The JES operation principle is based on a fully superconducting one-dimensional (1D) Josephson junction (JJ), i.e.,  two superconducting leads coupled by a superconducting nanowire with lateral dimensions smaller than its coherence length ($\xi$).   
The transition to the dissipative state can be understood to be due to $2\pi$ phase-slips, qualitatively similar to the tilted washboard potential model (WP) of JJs, where a phase particle moves in the WP under action of friction forces \cite{Barone1982,Tinkham1996}. 
The effective WP profile strongly depends on both bias current ($I$) through the junction and Josephson energy ($E_J$) \cite{Bezryadin2012} [see left panels of Figs. \ref{fig:Fig1}(b)-\ref{fig:Fig1}(c)]. 
In particular, for a 1D nanowire JJ the escape barrier can be written as \cite{Bezryadin2012}
\begin{equation}
U(I,E_J)\sim 2 E_J\left(1-I/I_C \right)^{5/4}\text{,}
\label{eq:potential}
\end{equation}
where $E_J=\Phi_0I_C/2\pi$, $\Phi_0\simeq 2.067\times 10^{-15}$ Wb is the flux quantum, and $I_C$ is the JJ critical current. 
Equation \ref{eq:potential} shows that the phase particle escape from a potential minimum (and, therefore, the corresponding transition of the junction to the dissipative state) can be finely controlled either by rising $I$ or by suppressing $E_J$.

Thanks to the exploitation of a 1D JJ,  the JES benefits from a two-fold advantage. 
On the one hand, transverse dimensions smaller than $\xi$ ensure a constant superconducting wave function along the wire cross section leading to uniform superconducting properties. 
On the other hand, a nanowire width ($w$) much smaller than the London penetration depth ($\lambda_L$) guarantees the supercurrent density in the JJ to be homogeneous when current-biasing the JES, and a uniform penetration of the film by an external magnetic field.

For comparison, in a sufficiently shunted overdamped limit, we can evaluate the resistance ($R$) versus temperature ($T$) characteristics of a JJ for different values of $I$ and $E_J$ by calculating the current derivative of the voltage drop \cite{Ivanchenko1969} 
\begin{equation}
V(I, E_J, T)=R_N \left( I - I_{C,0} \operatorname{Im}\frac{\mathcal{I}_{1-iz}\left( \frac{E_J}{k_BT}\right) }{\mathcal{I}_{-iz}\left( \frac{E_J}{k_BT}\right)}\right)\text{,}
\label{eq:voltage}
\end{equation}
where $I_{C,0}$ is the JJ zero-temperature critical current, $\mathcal{I}_{\mu}(x)$ is the modified Bessel function with imaginary argument $\mu$, and $z=\frac{E_J}{k_BT}\frac{I}{I_C}$. The details of the theoretical model are in the Supplemental Material \cite{SU}.

Specifically, an increase of $I$ leads to a sizable lowering of the resistive transition temperature accompanied by its \emph{narrowing}, as shown in the right panel of Fig. \ref{fig:Fig1}(b) for $E_J=E_{J0}$ (where $E_{J0}$ is the zero-temperature Josephson energy).
Similarly, the transition temperature can be reduced by decreasing $E_J$, but its width results wider than in the presence of a sizable bias current flowing through the nanowire [see the right panel of Fig. \ref{fig:Fig1}(c)].
In fact, the black dotted curves are calculated for $E_J=E_{J0}$ by varying $I$ in order to have the same transition temperature ($T_C$) given by the reduction of Josephson energy.
The temperature response of $R$  originating from the biasing current yields transitions which are \emph{sharper} than those obtained by simply reducing $E_J$.

Note that $E_J$ can be suppressed, for instance, by applying an external magnetic field ($B$). 
However, the use of $B$ can be detrimental for several applications, since it typically broadens the superconducting transition \cite{Zant1992}. 
As we shall show, tuning the nanowire JJ through current injection will prove to be an excellent strategy to achieve near-to-ideal nanosensors with ultimate performance.

\section{JES structure and basic characterization}

\begin{figure} [t!]
        \begin{center}
                \includegraphics [width=0.9\columnwidth]{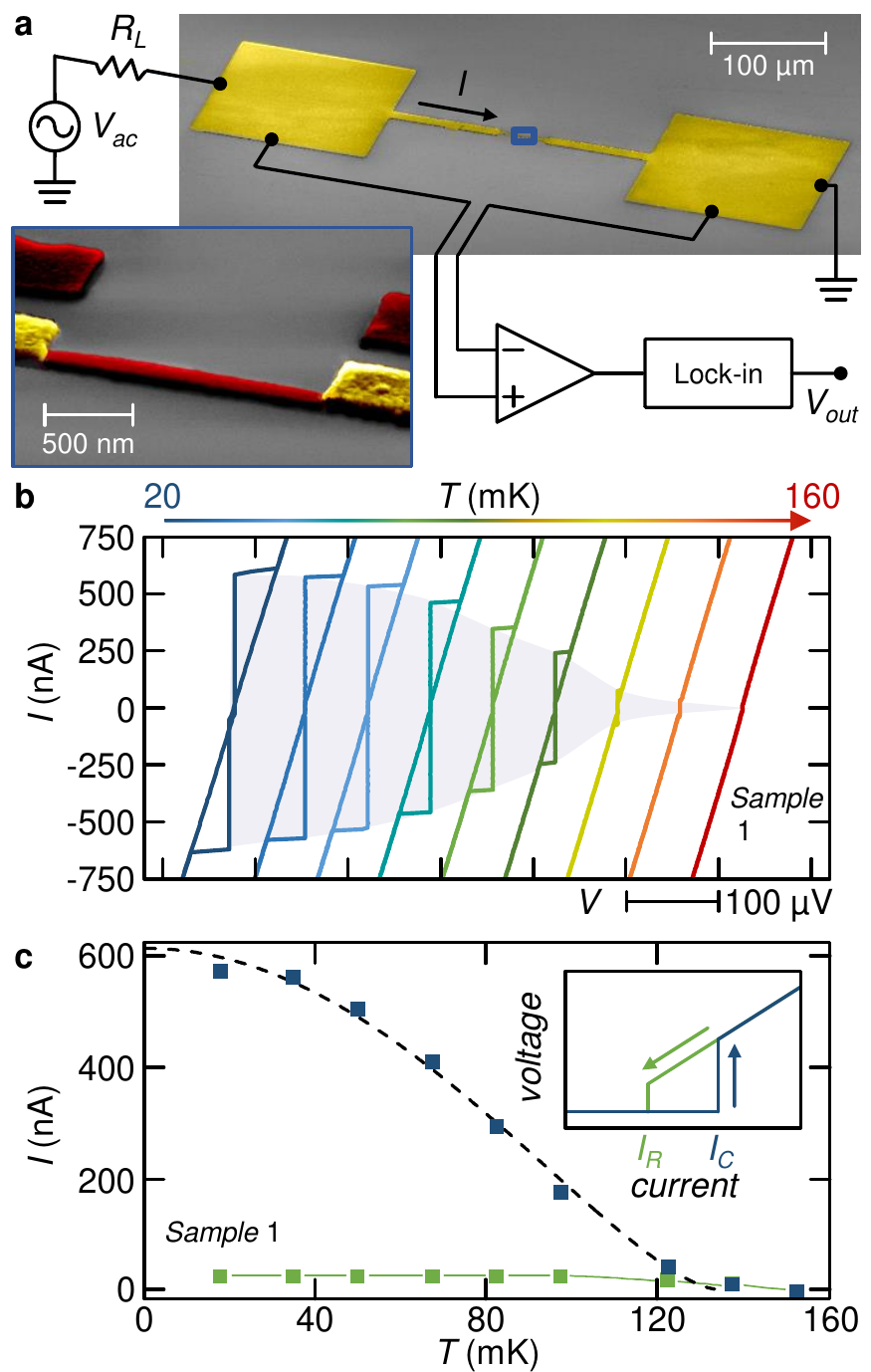}
        \end{center}
        \caption{\label{fig:Fig2}\textbf{Realization of the JES.} 
        \textbf{(a)} False-color scanning electron micrograph of a typical JES.
The nanosensor is AC current-biased (amplitude $I$) and the voltage drop across the wire ($V_{out}$) is measured via a voltage pre-amplifier connected to a lock-in amplifier. $R_L$ is a high-impedance load resistor.
 Inset: blow-up of the core of the JES showing the Al/Cu nanowire  (red) in contact with thick Al leads (yellow). 
\textbf{(b)} Back-and-forth DC current vs voltage ($I-V$) characteristics of a typical JES measured at different temperatures $T$ ranging from 20 mK to 160 mK in steps of $\sim20$ mK.
The curves are horizontally shifted for clarity. 
\textbf{(c)} Temperature evolution of the critical ($I_C$, blue) and retrapping ($I_R$, green) current. Dashed black line: critical current prediction of the Bardeen model \cite{Bardeen1962} (see Appendix A for details). Green line is a guide to the eye for the retrapping current.
Inset: the nanowire switches from the superconducting to the normal state at $I_C$, while the transition from the resistive to the dissipationless regime occurs at $I_R$. 
The hysteretic behavior stems from Joule heating when transitioning from the normal to the superconducting state \cite{Courtois2008}. 
$I_R$ is almost $T$-independent, and it is $\simeq26.6$ nA.}
\end{figure}

The realization of a typical JES is shown in the pseudo-color scanning electron micrograph displayed in Fig. \ref{fig:Fig2}(a). 
The JESs were fabricated by electron-beam lithography and two angles shadow-mask electron-beam evaporation of metals onto an oxidized silicon wafer through a suspended resist mask.  
The 1D sensor active region consists of a bilayer of Al ($t_{Al}=10.5$ nm) and Cu ($t_{Cu}=15$ nm) evaporated at an angle of 0$^{\circ}$. 
The total volume of the sensor active region is $\mathcal{V}_{w}=\mathcal{V}_{Al}+\mathcal{V}_{Cu}\simeq3.83\times 10^{-21}$ m$^{-3}$, with $\mathcal{V}_{Al}\simeq1.58\times 10^{-21}$ m$^{-3}$ and $\mathcal{V}_{Cu}\simeq2.25\times 10^{-21}$ m$^{-3}$.  
The lateral 40-nm-thick Al banks were then evaporated at an angle of 40$^{\circ}$.

All measurements were performed in a filtered He$^3$-He$^4$ dry dilution refrigerator at different bath temperatures in the range $20-160$ mK.
The resistance $R$ vs temperature characteristics of the JES and of the Al banks were obtained by conventional four-wire low-frequency lock-in technique at $13.33$ Hz.
To this end, AC excitation currents with typical root mean square amplitudes $I\simeq15-380$ nA were imposed through the device. 
The current was generated by applying an AC voltage bias ($V_{ac}$) to a load resistor of impedance ($R_L$) much larger than the sample resistance ($R_L=100$ k$\Omega\gg R$). 
The critical temperature of the Al banks was measured with the same set-up. 
The $I$ vs $V$ characteristics of the nanowires
were obtained by applying a low-noise DC biasing current, while the
voltage drop  was measured
via a room-temperature battery-powered differential preamplifier.

Josephson transport in the nanosensor is highlighted by the DC current ($I$) vs voltage ($V$) characteristics shown in Fig. \ref{fig:Fig2}(b) for bath temperatures ($T$) ranging from 20 mK to 160 mK. The wire normal-state resistance is $R_N\simeq 77$ $\Omega$, and the typical heating-induced hysteretic behavior of the $I-V$ curves is observed \cite{Courtois2008}. 
On the one hand, the critical current obtains a maximum $I_{C}\simeq 575$ nA at $T\simeq20$ mK [see Fig. \ref{fig:Fig2}(c)], and monotonically decreases with $T$ following the prediction of Bardeen \cite{Bardeen1962} (see Appendix A for details). On the other hand, the retrapping current ($I_R\simeq 26.6$ nA) is constant in the whole temperature range. 

\begin{figure*} [ht!]
        \begin{center}
                \includegraphics [width=1\textwidth]{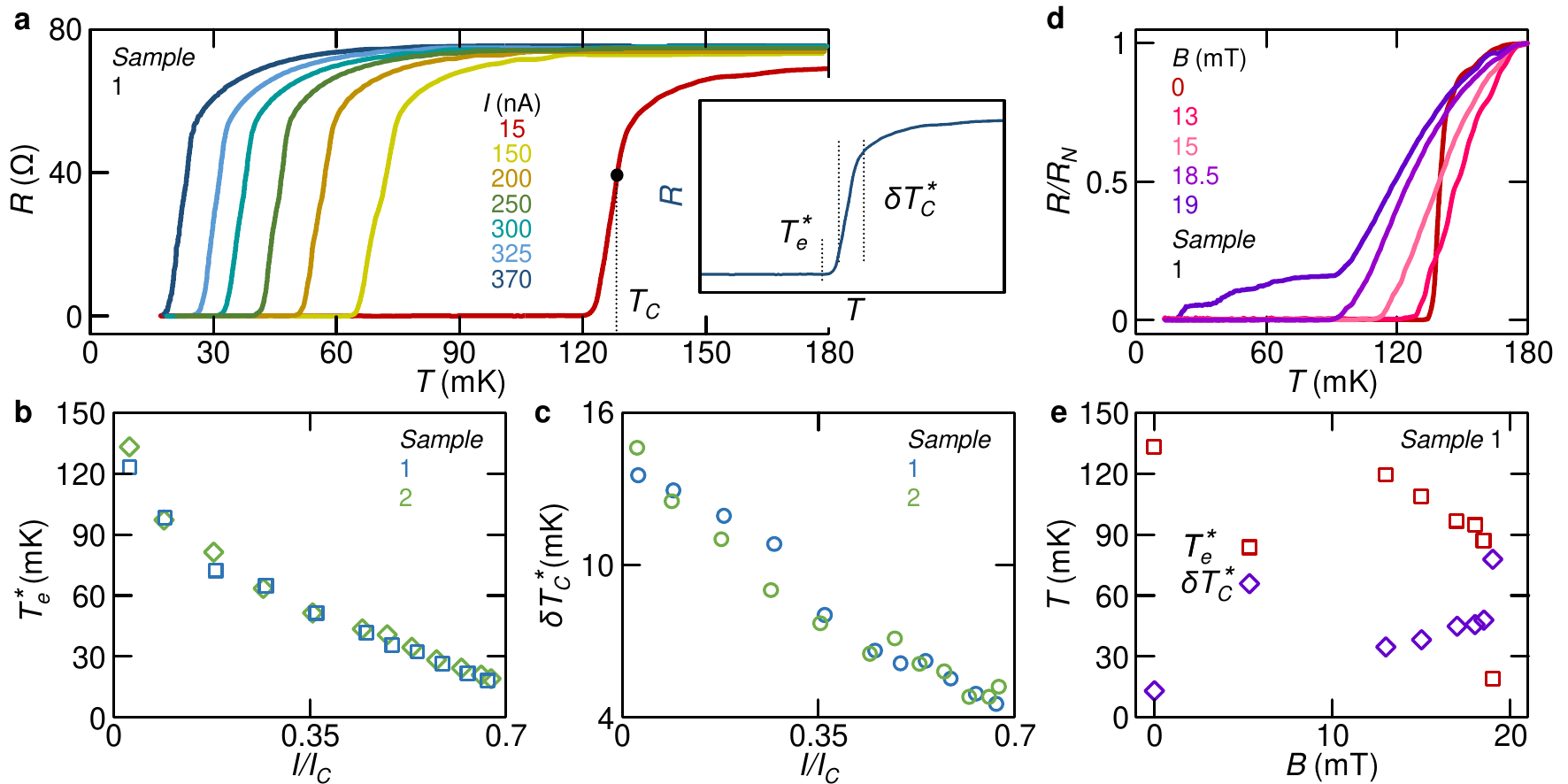}
        \end{center}
        \caption{\label{fig:Fig3}\textbf{Tuning the properties of the JES.} 
        \textbf{(a)} Selected resistance ($R$) vs temperature ($T$) characteristics  for different values of AC bias current amplitude $I$. 
        Inset: Sketch of the temperature dependence of the JES resistance. 
The current-dependent escape temperature $T_e^*$  as well as the phase transition width  $\delta T_C^*$ are indicated. 
$T_e^*$ monotonically decreases by increasing $I$ while the transition becomes \emph{sharper} by increasing the biasing current.
$T_C\simeq 130$ mK is the critical temperature of sample 1.
\textbf{(b)} Full behavior of $T_e^*$ vs $I$ for two different JESs. For large $I$, $T_e^*$ can be as small as $\simeq 20$ mK. 
\textbf{c}, Width of the phase transition $\delta T_C^*$ vs $I$ for two different JESs.  $\delta T_C^*$ is suppressed by a factor of 4 at the largest biasing currents. 
Note the fine tunability of $T_e^*$ provided by the injection current.
\textbf{(d)}  $R$ vs $T$ characteristics for different values of the perpendicular-to-plane magnetic field ($B$). 
The sizable widening of the phase transition is likely to stem from depairing in the nanowire induced by $B$ \cite{Zant1992}. 
\textbf{(e)} $T_e^*$ and $\delta T_C^*$  vs $B$. 
$T_e^*$ shifts towards lower values by rising $B$, but the transition becomes much broader at higher fields. 
Yet, $T_e^*$ is hardly tunable at large values of $B$.}
\end{figure*}

As stated, the JES working principle is based on a 1D nanowire JJ.  Indeed, the coherence length ($\xi\simeq 220$ nm) and the London penetration depth ($\lambda_L\simeq 970$ nm) of the nanowire are much larger than the wire width ($w=100$ nm) and total film thickness ($t_{w}=25.5$ nm) thereby providing the frame of a 1D junction (see Appendix A). 
Yet, the nanowire length ($\sim 6.8\; \xi$) reduces the influence of the superconducting proximity effect arising from the clean contact with the lateral Al leads \cite{Tinkham1996}. 
We also note that the maximum magnetic field created by the critical current flow, $B_{max}\simeq 5$ $\mu$T, is negligibly small compared to the out-of-plane critical magnetic field of the wire ($B_C\simeq 21$ mT) thus implying a vanishing effect on the JES.

\begin{figure*} [t!]
        \begin{center}
                \includegraphics [width=1\textwidth]{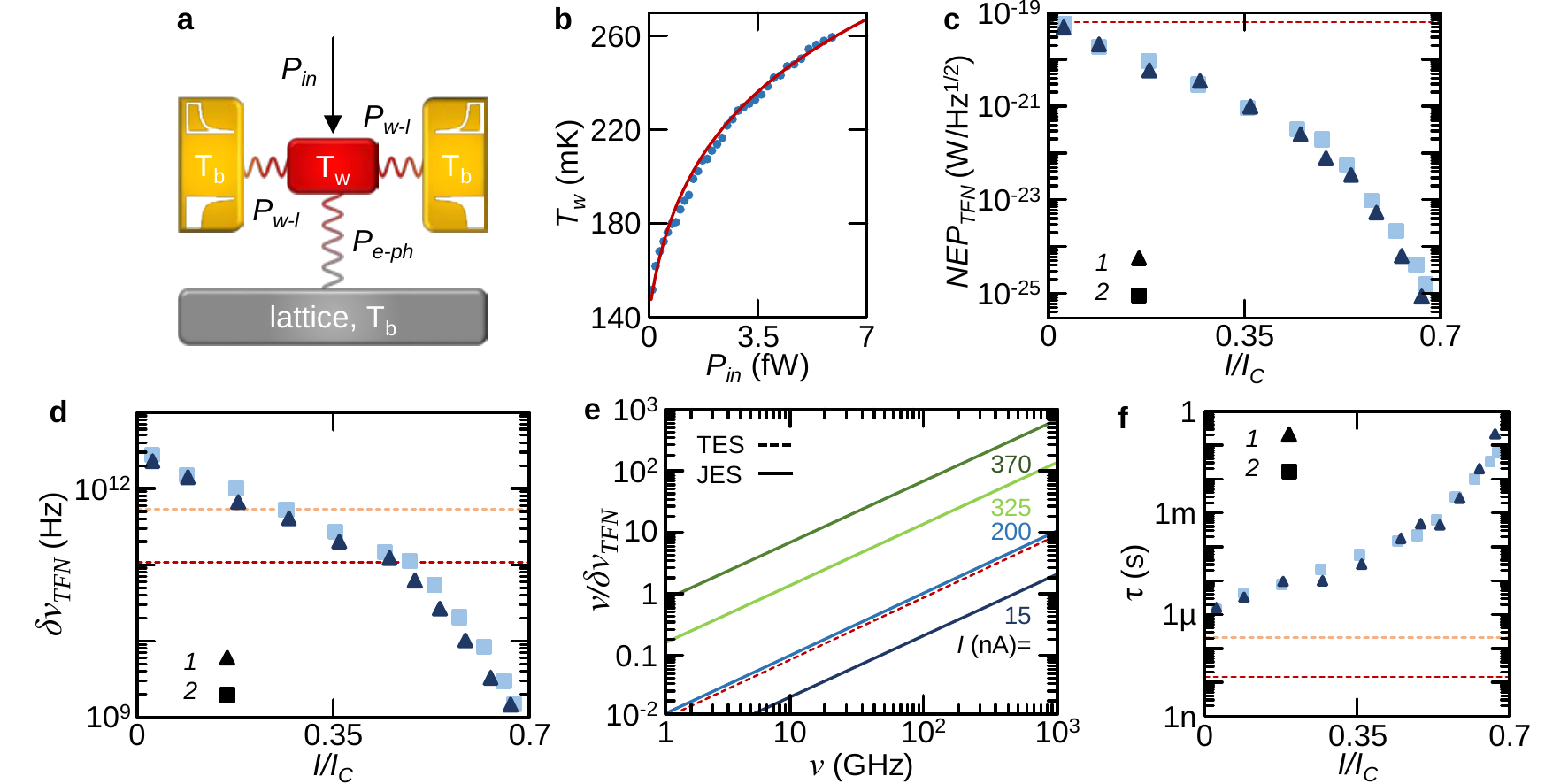}
        \end{center}
        \caption{\label{fig:Fig4}\textbf{Performance of the JES.} 
        \textbf{(a)} Thermal model highlighting the predominant heat exchange mechanisms occurring in the nanosensor. 
        $P_{in}$ is the power coming from the incident radiation, $P_{e-ph}$ is the heat exchanged between electrons  in the nanowire residing at temperature $T_w$ (red box) and lattice phonons residing at bath temperature $T_b$ (gray box), and $P_{w-l}$ is the electron heat current flowing from the nanowire to the superconducting leads residing at $T_b$ (yellow boxes). 
        \textbf{(b)} Electronic temperature in the nanowire $T_w$ (blue dots) vs $P_{in}$ recorded at $T_b=147.5$ mK. The red line represents the theoretical behavior %(see SI).
        (see Supplemental Material for details \cite{SU}).
        \textbf{(c)} Deduced NEP$_{TFN}$ vs $I$ for sample $1$ (triangles) and $2$ (squares).
        \textbf{(d)} Frequency resolution $\delta \nu_{\rm TFN}$  vs $I$ for sample $1$ (triangles) and $2$ (squares).
        \textbf{(e)} Resolving power $\nu/ \delta \nu_{\rm TFN}$ vs  frequency $\nu$ calculated for sample $1$. $\nu/ \delta \nu_{\rm TFN}$ increases by rising the bias current (from blue to green). 
        \textbf{(f)} Time constant $\tau$ vs $I$ for sample 1 (triangles) and $2$ (squares). Dashed lines in panels c-f indicate the expected figures of merit when the sensor is operated at $T_C$ in TES mode for sample 1 (red) and sample 2 (yellow).
      }
    \end{figure*}

\section{Current modulation of the R vs T characteristics}

We investigated the behavior of the JES by recording the resistance  versus temperature characteristics for several amplitudes $I$ of low-frequency AC bias current [see Fig. \ref{fig:Fig3}(a)]. 
The $R(T)$ characteristics monotonically shift towards low temperatures by increasing $I$, almost preserving the same shape up to the largest current amplitude. 
In particular, $I$ was varied between $\sim 3\%$ and $\sim 64\%$ of $I_C$.
Note that although the transition curves shift towards low $T$ by increasing $I$, the nanowire electronic temperature $T_w$ in the middle of the transition under current injection is not expected to coincide anymore  with the bath temperature $T_b$. 
Indeed, when transitioning to the normal state,
electrons in the nanowire are Joule overheated with respect to $T_b$ by the bias current (with final $T_w\lesssim T_C$), thus preventing the operation of the nanosensor as a conventional TES biased at those low bath temperatures, without additional shunting. 
By contrast, when operated in the dissipationless regime, i.e., as an \emph{escape} sensor,   $T_w$ coincides with $T_b$.

From the $R$ vs $T$ curves we can specify a current-dependent temperature related to the resistive transition, i.e.,  
the \emph{escape} temperature [$T^*_e(I)$]. The latter is the maximum value of $T$ providing a \emph{zero} nanowire resistance  (see the inset of Fig. \ref{fig:Fig3}(a) and Appendix A).
The $T^*_e(I)$ characteristics for two JES samples  are shown in Fig. \ref{fig:Fig3}(b). 
In particular, $T^*_e$ is monotonically reduced by increasing $I$ with a minimum value $\sim20$ mK for $I=370$ nA corresponding to $\sim 15\%$ of the nanowire intrinsic critical temperature, $T_C\sim 130$ mK. Moreover, the transition width ($\delta T_C^*$) \emph{narrows} by increasing $I$ [see Fig. \ref{fig:Fig3}(c)]. In particular, $\delta T_C^*$ is suppressed by a factor of $4$ at the largest current amplitude, mirroring the expected changes in the switching as shown in Fig. \ref{fig:Fig1}(b). 

In addition, to prove the complementary tuning of the WP
through the suppression of $I_C$ (see Eq. \ref{eq:potential}), we applied a perpendicular-to-plane magnetic field. 
The resulting shape of the corresponding transition degrades dramatically in the presence of $B$ [see Fig. \ref{fig:Fig3}(d)]. 
n particular, the $R$ vs $T$ characteristics appear to be scarcely tunable, while the onset of the transition is almost unaffected. The extreme broadening of the superconducting to dissipative state transition can be explained in terms of flux penetration in the nanowire and in the aluminum banks, since our nanowires are thinner than the London penetration depth for magnetic field (see Appendix A for details).
%Thus we can only define $T_e^*$, since the effective critical temperature loses significance when the transition is so broad.% 
$T_e^*$ shows a stark variation at values of $B\to B_C$,
and is joined to the outbreak of the transition width, as displayed in Fig. \ref{fig:Fig3}(e). 
The above results in a finite magnetic field validate therefore the bias current as an ideal tool to control the JES properties. 

\section{Performance of JES-based bolometers and calorimeters}

Insight into the behavior of the JES can be gained by considering the predominant heat exchange mechanisms occurring in the nanodevice, as schematically depicted in the thermal model of Fig. \ref{fig:Fig4}(a). 
The absorption of external incident radiation ($P_{in}$)  leads to the increase of the nanowire electronic temperature $T_w$. Yet, the two lateral superconducting Al leads (residing at bath temperature $T_b$) serve as Andreev mirrors \cite{Andreev1964} thereby suppressing heat out-diffusion ($P_{w-l}$) from the nanowire. 
As a consequence, the main thermal relaxation channel in the system stems from heat exchange with lattice phonons ($P_{e-ph}$) residing at $T_b$.
For a normal metallic thin film, $P_{e-ph,n}=\Sigma_w \mathcal{V}_{w}\left(T_w^5-T_b^5 \right)$ \cite{Irwin1995,Giazotto2006}, where $\mathcal{V}_w$ is the nanowire volume, and $\Sigma_w$ is the electron-phonon coupling constant of the bilayer. 
$\Sigma_w$ was determined through energy-relaxation experiments \cite{Giazotto2006} by injecting a known power, and by measuring the resulting steady-state electron temperature established in an \emph{ad hoc} fabricated identical wire  kept  above its critical temperature.
Figure \ref{fig:Fig4}(b) shows the $T_w$ vs $P_{in}$ characteristic (blue dots) recorded at $T_{b}=147.5$ mK along with a fit to the data which allows to extract the electron-phonon coupling constant in the Al/Cu nanowire, $\Sigma_w \simeq 1.15 \times 10^9$ Wm$^{-3}$K$^{-5}$ (see Supplemental Material \cite{SU}).

Yet, since the JES is operated in the superconducting state at $T_e^*(I)$, the latter can be substantially smaller than $T_C$ depending on the current amplitude. 
At sufficiently low temperature, the electron-phonon heat exchange in a superconductor is \emph{exponentially} suppressed with respect to the normal state owing to the presence of the energy gap, i.e., $P_{e-ph,s} \propto P_{e-ph,n} \exp{[-\Delta_w/(k_BT_e^*)]}$ \cite{Timofeev2009}, where $\Delta_w\simeq 23\,\mu$eV is the pairing potential in the nanowire (see Supplemental Material for details).
As we shall argue, the operation deeply in the superconducting state dramatically improves the JES key figures of merit for radiation detection.

In general, the performance of a bolometer can be quantified by the NEP that is the input power resolution per unit bandwidth. For the JES, the NEP is bounded by thermal fluctuations  between the electron and phonon system in the nanowire \cite{Giazotto2006}. Other limitations to the resolution can arise from the switching measurement, which we assume is optimized to be sub-dominant.
The thermal fluctuation noise (TFN)-limited NEP$_{TFN}$ can be extracted by substituting the measured parameters of the JES, such as $I_C$ vs. $T$, $R(T)$ characteristics, $\Delta_w$ and $\Sigma_w$, in the widespread and well known equations for superconducting radiation sensors \cite{Giazotto2006} (see also Supplemental Material \cite{SU}).
The NEP$_{TFN}$ monotonically decreases by increasing the current amplitude, and turns out to be 
 \textit{in-situ} finely controlled by tuning $I$ [see Fig. \ref{fig:Fig4}(c)]. 
In particular, the JES points to  noise values  which are several orders of magnitude smaller than so far reported. 
Specifically, the best extracted NEP$_{TFN}$  obtains values as low as $\sim 1\times 10^{-25}$ W/$\sqrt{\text{Hz}}$ for  $I=370$ nA at $\sim 18$ mK. 
By contrast, in the normal state, i.e., when it is operated as a conventional TES, the sensor is expected to provide a much higher NEP$_{TFN}\sim 6\times 10^{-20}$ W/$\sqrt{\text{Hz}}$, because the electron-phonon thermalisation in the active region is stronger.

In pulsed detection mode,
a relevant figure of merit of a radiation sensor is represented by the frequency resolution ($\delta \nu$), i.e., the minimum detectable energy for an incident single photon 
(see Supplemental Material for details \cite{SU}).
We here assume the JES measurement is performed accurately and sufficiently fast \cite{Zgirski2018,Wang2018} compared to the thermal relaxation time $\tau$, and estimate the limitation from the TFN.
Figure \ref{fig:Fig4}(d) emphasizes the strong dependence of $\delta \nu_{\rm TFN}$ on $I$ which displays variations over $3$ orders of magnitude. 
For $\delta \nu_{\rm TFN}$, this limit can be as low as $\sim 2$ GHz at $370$ nA, and would enable single-photon sensing at unprecedented low energies. 
When operating the nanosensor as TES, 
$\delta \nu$ is roughly two orders of magnitude larger, and obtains $\sim 100$ GHz.
The limitation to single-photon sensing capability from TFN in our JES is highlighted by  the resolving power ($\nu/ \delta \nu_{\rm TFN}$), calculated vs incident radiation frequency in Fig. \ref{fig:Fig4}(e). 
$\nu/ \delta \nu_{\rm TFN}$ can reach $\sim 40$ at $50$ GHz, and  $ \sim 800$ at $1$ THz both for $370$ nA, whereas it can be   $\sim 10$ at $1$ THz when operating the sensor as a TES. Since the power associated to 1 THz photons is lower than $1\times 10^{-18}$ W and the power dissipated through the phonons at $T=260$ mK is about 6 fW [see Fig. \ref{fig:Fig4}(b)], the electronic temperature in the wire is always lower than $0.3 T_{C,Al}$. Therefore, we can exclude thermal diffusion through the aluminum banks \cite{Andreev1964} thus considering the active region fully thermally isolated from the electrodes.

% , whereas it can be   $\sim 10$ at $1$ THz when operating the sensor  in  TES-mode at $T_C$.

We wish to finally comment onto the JES time constant ($\tau$), which is one of the fundamental figures of merit for a radiation sensor. It is basically given by the ratio between the electron heat capacitance and the electron-phonon heat conductance in the nanowire \cite{Giazotto2006}, since heat conduction through the lateral Al electrodes is negligible in a JES. 
In pulsed detection mode $\tau$ determines the minimum speed of the read-out electronics (which has to be faster than $\tau$), and the minimum time separation for the independent detection of two photons. 
Figure \ref{fig:Fig4}(f) shows the expected JES time constant vs bias current $I$, as deduced from the experimental data 
(see Supplemental Material for details).
In particular, $\tau$ increases monotonically by increasing $I$, and  varies between $\sim 1\,\mu$s at low current amplitude and  $\sim 100\,$ms at $370$ nA.
In general, when used at $T_C$ as a TES, the nanosensor can provide a much faster response  than the JES in the whole bias current range ($\tau \sim 10...100$ ns) thanks to the higher operation temperature and electron-phonon thermal relaxation.

\section{Conclusions}

We have conceived and demonstrated an innovative  hypersensitive superconducting radiation sensing element supplied with the capability of \textit{in-situ} fine tuning its performances by a current bias. 
Our nanosensor has the potential to drive radiation detection in the gigahertz regime towards unexplored levels of sensitivity  
%with respect to existing sensing architectures \cite{Kokkoniemi2019,Chen2011,Nararajan2012,Karimi2020,Virtanen2018,Giazotto2008, Kuzmin2019,Govenius2016}
by lowering the thermal fluctuation limitation to NEP down to $\sim 1\times 10^{-25}$ W/$\sqrt{\text{Hz}}$, with a corresponding limit in frequency resolution at $\sim 2$ GHz. 
The JES is expected to have significant impact in radio astronomy \cite{Rowan2009,Marcote2020,Tabatabaei2017,Sironi1999,Falchi1988,Issaoun2019,Adams2020,Ade2019}, space spectroscopy \cite{Odea1998} and dark matter search \cite{Capparelli2016,Irastorza2018}, since its working mechanism could allow, in principle,  the immediate replacement of TESs in already existing experiments and telescopes.
Furthermore, the JES could have countless applications in several fields of quantum technology where extrasensitive photon detection is a fundamental task, such as sub-terahertz communication \cite{Federici2010}, quantum computation \cite{Obrien2007} and quantum cryptography \cite{Gisin2002,Tittel2019}. 

\section*{Acknowledgements}
We acknowledge F. S. Bergeret, G. De Simoni, E. Strambini, A. Tartari, and G. Marchegiani for fruitful discussions.
The authors acknowledge  the European Union's Horizon 2020 research and innovation programme under the grant No. 777222 ATTRACT (Project T-CONVERSE) and under grant agreement No. 800923-SUPERTED. The authors acknowledge CSN V of INFN under the technology innovation grant SIMP. The work of F.P. was partially supported by the Tuscany Government (Grant No. POR FSE 2014-2020) through the INFN-RT2 172800 project. The work of V.B. is partially funded by the European Union (Grant No. 777222 ATTRACT) through the T-CONVERSE project.

\section*{APPENDIX A: Device parameters}
The temperature dependence of the critical current of the nanowire can be fitted through the phenomenological equation \cite{Bardeen1962} $I_C(T)=I_{C,0}\left[ 1-\left( T/T_C\right)^2\right] ^{3/2}$, where $I_{C,0}$ is the zero-temperature critical current. The fit provides $I_{C,0}\simeq615$ nA and a critical temperature $T_{C,fit}\simeq 133$ mK, which is good in agreement with the experimental value obtained from the resistance versus temperature characteristics.

The Al/Cu bilayer can be considered as an uniform superconductor, since it lies within the Cooper limit \cite{DeGennes1964,Kogan1982}. Here, we assume a transparent Al/Cu interface. Moreover, the aluminum layer respects $t_{Al}=10.5\;\text{nm}\ll\xi_{Al}=\sqrt{\hbar D_{Al}/\Delta_{Al}}\simeq80$ nm (with $\hbar$ the reduced Planck constant, $D_{Al}=2.25\times 10^{-3}$ m$^2$s$^{-1}$ is the diffusion constant of Al and $\Delta_{Al}\simeq200\;\mu$eV its superconducting energy gap), while the copper film obeys to $t_{Cu}=15\;\text{nm}\ll\xi_{Cu}=\sqrt{\hbar D_{Cu}/(2\pi k_B T)} \simeq255$ nm (with $D_{Cu}=8\times 10^{-3}$ m$^2$s$^{-1}$ the copper diffusion constant and the temperature is chosen to the worst case scenario $T=150$ mK).

From the nanowire normal-state resistance ($R_N\simeq 77$ $\Omega$) we determined the superconducting coherence length in the active region $\xi=\sqrt{l \hbar/[(t_{Al}N_{Al}+t_{Cu}N_{Cu}) R_Ne^2\Delta_{w}]}\simeq 220$ nm, where $e$ is the electron charge, while $N_{Al}=2.15\times 10^{47}$ J$^{-1}$m$^{-3}$ and $N_{Cu}=1.56\times 10^{47}$ J$^{-1}$m$^{-3}$ are the density of states at the Fermi level of aluminum and copper, respectively. 
As a consequence, the nanowire obeys to $t_w=t_{Al}+t_{Cu}=25.5\;\text{nm}\ll\xi$ thus ensuring constant pairing potential along the out-of-plane axis.
The superconducting energy gap of the bilayer, $\Delta_{w}=23$ $\mu$eV, has been determined by tunnel spectroscopy performed on \textit{ad hoc} fabricated nominally identical wires equipped with Al tunnel probes.
For further details, see  Supplemental Material.
The London penetration depth was determined as $\lambda_L=\sqrt{\hbar (t_{Al}+t_{Cu})w R_N/(\pi \mu_0 l \Delta_{w})}\simeq 970$ nm, where $\mu_0$ is the magnetic permeability of vacuum.
The maximum magnetic field generated by the bias current at the wire surface reads $B_{I,max}=\mu_0 I_{C,0}/(2\pi t_{w})\simeq 5$ $\mu$T, where $I_{C,0}$ is the zero-temperature critical current, and $t_w=t_{Al}+t_{Cu}$ is the total thickness of the JES active region.
Finally, the critical temperature of the Al banks was $T_{C,Al}\simeq1.3$ K.

We would like to note that the energy gap expected for a non-superconducting Al/Cu bilayer would e $E_g\simeq3\hbar D_w/L^2\simeq5\;\mu$eV, where $D_w=(D_{Al}t_{Al}+D_{Cu}t_{Cu})/t_{w}\simeq5.6\times 10^{-3}$ m$^2$s$^{-1}$ is the wire diffusion coefficient and $L=1.5\;\mu$m is its length. This values is less than 1/4 of the measured gap. Therefore, the intrinsic superconductivity of the bilayer dominates.  

The current-dependent escape temperature [$T_e^*(I)$] is defined as the maximum value of temperature providing $R(I)=0$, i.e., when the JES is in the dissipationless state.

\section*{SUPPLEMENTARY INFORMATION}
\section{Model of the tunable Josephson escape sensor}
The Josephson escape sensor (JES) is composed of two superconducting leads interrupted by a one dimensional superconducting nanowire (both length and width are shorter than its superconducting coherence length $\xi$ and its London penetration depth $\lambda_L$), that is a 1D fully superconducting Josephson junction (JJ). The electronic transport of such a system can be described through the overdamped resistively shunted junction
(RSJ) model \cite{tinkham}. Here, the transition to the dissipative state is attributed to a $2\pi$ phase-slip of a phase particle moving in a tilted washboard potential (WP) under the action of a friction force. Within the RSJ model, the dependence on the bias current ($I$) of the stochastic phase difference [$\varphi(t)$] over the JJ is given by
\begin{equation}
\frac{2e}{\hbar} \frac{\dot{\varphi}(t)}{R_N} + I_C \sin{\varphi(t)} = I + \delta I_{th}(t) \text{,}
\label{CPR}
\end{equation}
where $\hbar$ is the reduced Planck constant, $I_C$ is the junction critical current and $\left\langle \delta I_{th}(t)\delta I_{th}(t')\right\rangle=\frac{K_BT}{R_N}\delta{(t-t')}$ is the thermal noise generated by the shunt resistor $R_N$, with $k_B$ the Boltzmann constant and $T$ the temperature.

\subsection{General solution}
The 1D nature of the JJ composing the JES entails the homogeneous flow of the bias current across the nanowire section. Under the assumption of phase-slips induced only from sources outside the junction, the voltage drop across the JJ reads \cite{Ivanchenko1969}
\begin{equation}
V(I, E_J, T)=R_N \left( I - I_{C,0} \operatorname{Im}\frac{\mathcal{I}_{1-iz}\left( \frac{E_J}{k_BT}\right) }{\mathcal{I}_{-iz}
\left( \frac{E_J}{k_BT}\right)}\right)\text{,}
\label{eq:voltage}
\end{equation}
where $I_{C,0}$ is the JJ zero-temperature critical current, $\mathcal{I}_{\mu}(x)$ is the modified Bessel function with imaginary argument $\mu$, $E_J=\Phi_0I_C/2\pi$ (with $\Phi_0\simeq 2.067\times 10^{-15}$ Wb the flux quantum) is the Josephson energy, and $z=\frac{E_J}{k_BT}\frac{I}{I_C}$. Thus, the current and Josephson energy dependent $R(T)$ characteristics of the JES can be calculated by
\begin{equation}
R(I, E_J, T)=\frac{\text{d}V(I, E_J, T)}{\text{d}T}\text{.}
\label{resistance}
\end{equation}

In the limit of $I=0$ the junction resistance can be simplified as \cite{tinkham}
\begin{equation}
R(I=0, E_J, T)= R_N \frac{1}{\mathcal{I}_{0}(\frac{E_J}{k_BT})^2} \text{,}
\label{resistanceTink}
\end{equation}
where $\mathcal{I}_{0}$ is the zero-order modified Bessel function.

\subsection{Simplified solutions for $I\to0$ and $I\to I_C$} \label{Ivan}
A simplified picture of the dependence of $R(T)$ on bias current and Josephson energy can be provided by considering the influence of $I$ and $E_J$ on the WP. In fact, the escape barrier for a JJ takes the form \cite{Bezryadin2012}
\begin{equation}
U(I,E_J)\sim 2E_J\left(1-I/I_C \right)^{5/4}\text{,}
\label{eq:potential}
\end{equation}
where the exponent $5/4$ stems from the nanowire nature of the constriction. Equation \ref{eq:potential} shows that $I$ and $E_J$ control the escape of a phase particle from a potential minimum of the WP. The resulting thermal fluctuation induced voltage at low temperature reads \cite{Ambegaokar1969}
\begin{equation}
\begin{split}
V(I, E_J, T)\sim \\
R_NI_Ca(I/I_C)e^{-U(I,E_J)/(k_BT)} 
\Bigl(1 - e^{-\pi \hbar I/(ek_BT)}\Bigr)\,.
\end{split}
\label{voltageAmbe}
\end{equation}
where $a(x)\sim1$.

From this, we get the simplified results at low bias current ($I\to 0$)
\begin{equation}
\begin{split}
R(I\to 0, E_J,T)\sim \\
R_N \pi \frac{U(I\to 0, E_J)}{k_BT}e^{-U(I \to 0,E_J)/(k_BT)}\text{,}
\end{split}
\label{reszero}
\end{equation}
whereas at high bias current ($I\to I_C$)
\begin{equation}
\begin{split}
  R(I, E_J,T)\sim \\
  R_N f(I/I_C) \pi \frac{U(I, E_J)}{k_BT} e^{-U(I,E_J)/(k_BT)}
  \,,
  \end{split}
\label{resIc}
\end{equation}
where $f\propto -U'(I,E_J)/U(I,E_J) \propto I_C/(I_C - I)$.

The increase of $I$ and reduction of $E_J$ have similar effects on $U(I,E_J)$ (see Eq. \ref{eq:potential}), therefore the exponentials in Eqs. \ref{reszero} and \ref{resIc} produce comparable consequences to the $R(T)$ for $I\to I_C$ and $E_J\to 0$. On the contrary, we observe that the $R(T)$ rises faster for $I\to I_C$, because of the divergent prefactor $f(I/I_C)$ (see Eq. \ref{resIc}). The latter arises from the dependence of the potential barrier [$U(I,E_J)$] on the WP tilt. On the one hand, $I\to I_C$ and $E_J \to 0$ decrease the superconducting-to-dissipative transition temperature similarly to $E_J$ reduction. On the other hand, the increase of bias current improves the sharpness of the transition more than the suppression of Josephson energy. Therefore, we can conclude that bias current injection is the best strategy to reach ultimate sensing performance from a JES.

\begin{figure*}[t!]
        \centering
        \includegraphics[width=1\textwidth]{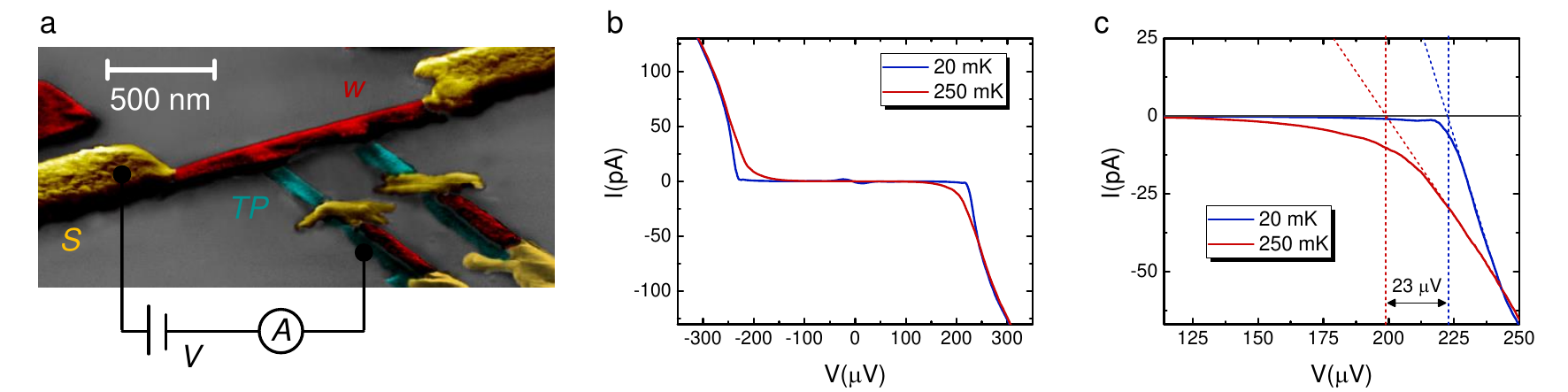}
        \caption{\textbf{Measurement of the superconducting energy gap of the nanowire.} (a) False-color scanning electron micrograph of the AD used to determine $\Delta_{w}$: Al banks (yellow) and probes (blue) are made of Al and the wire active region (red) is composed of an Al/Cu bilayer. The experimental setup used to obtain the $I-V$ characteristics is shown as well: a voltage ($V$) is applied to a couple probe-bank and the circulating current is measured. (b) $I-V$ characteristics recorded at T$_{b}$=20 mK (blue line) and T$_{b}$=250 mK (red line): the typical non-dissipative flat region and the switching points to the resistive-state are clearly visible. (c) $I-V$ characteristics zoomed around the positive voltage switching point: the difference between the projections of the linear part of the curves on the gray line ($I=0$) provides the energy gap of the bilayer $\Delta_w= 23\pm3\; \mu$eV.}
        \label{delta}
\end{figure*}

\section{Properties of the Al/Cu bilayer}
In order to provide a full description of the characteristics of the JES active region, we measured the energy gap ($\Delta_{w}$) and the electron-phonon energy relaxation constant ($\Sigma_w$) of the Al/Cu bilayer. 

\subsection{Auxiliary device design and fabrication}
\label{sec_fabrication}
The false-color scanning electron micrograph of the auxiliary devices (ADs) is shown in Fig. \ref{delta}-a. A typical AD consists of the same Al/Cu bilayer ($w$, red) of the JES contacted by two Al banks ($S$, yellow). In addition, the device is equipped with two Al tunnel probes ($TP$, blue) directly coupled to the wire allowing to perform both spectroscopy (for the measurement of $\Delta_{w}$) and thermometry (for the determination of $\Sigma_w$). 

The ADs were fabricated by electron-beam lithography and 3-angles shadow mask evaporation of metals onto a silicon wafer covered with 300 nm SiO$_{2}$ carried out in a ultra-high vacuum electron-beam evaporator. 
The 13-nm-thick Al probes were evaporated at an angle of -40$^{\circ}$ and then oxidized by exposition to 200 mtorr of O$_2$ for 5 minutes (we call the \textit{AlOx} layer as $I$). The Al/Cu bilayer ($t_{Al}=10.5$ and $t_{Cu}=15$ nm) was evaporated at an angle of 0$^{\circ}$. Finally, the 40-nm-thick Al electrodes were evaporated at an angle of 40$^{\circ}$. 

\subsection{Measurement of the energy gap $\Delta_{w}$}
The energy gap of Al/Cu bilayer was measured by considering the temperature dependence of the current-voltage ($I-V$) characteristics of the $S$-$w$-$I$-$TP$ tunnel Josephson junction (JJ), as sketched in Fig. \ref{delta}-a. Within this configuration, a quasiparticle tunneling current is seen for $V>(\Delta_{w}+\Delta_{TP})/e$ \cite{giazotto}, where $\Delta_{TP}$ is the energy gap of $TP$, and $e$ is the electron charge. Fig. \ref{delta}-b, shows the $I-V$ characteristics measured in the voltage range $\pm$350 $\mu$V at two different bath temperatures, T$_{b}$=20 mK (blue line) and T$_{b}$=250 mK (red line). As expected, by rising the temperature the critical voltage for switching to the resistive state is decreased.

In order to quantify $\Delta_w$, we zoomed the $I-V$ characteristics around the switching point at positive voltage bias (see Fig. \ref{delta}-c). The difference between the blue and the red line is due to the full suppression of the energy gap of the Al/Cu bilayer. In fact, $\Delta_{TP}(T)=\Delta_{0,TP}$ (where $\Delta_{0,TP}$ is the zero-temperature energy gap of the tunnel probe) in the complete temperature range, since in the worst case we have $T_{b}=250$ mK $<0.4T_{C,Al}$ \cite{tinkham}, where $T_{C,Al}\simeq1.3$ K is the measured critical temperature of the Al probes. Therefore, the difference of the projections of the linear part of the $I-V$ curves (to avoid the quasiparticle sub-gap conduction) at $I=0$ represents exactly $\Delta_{w}$. By repeating the measurements ten times, we obtained $\Delta_{w}=23\pm3$ $\mu$eV.

\begin{figure*}[t!]
        \centering
        \includegraphics[width=1\textwidth]{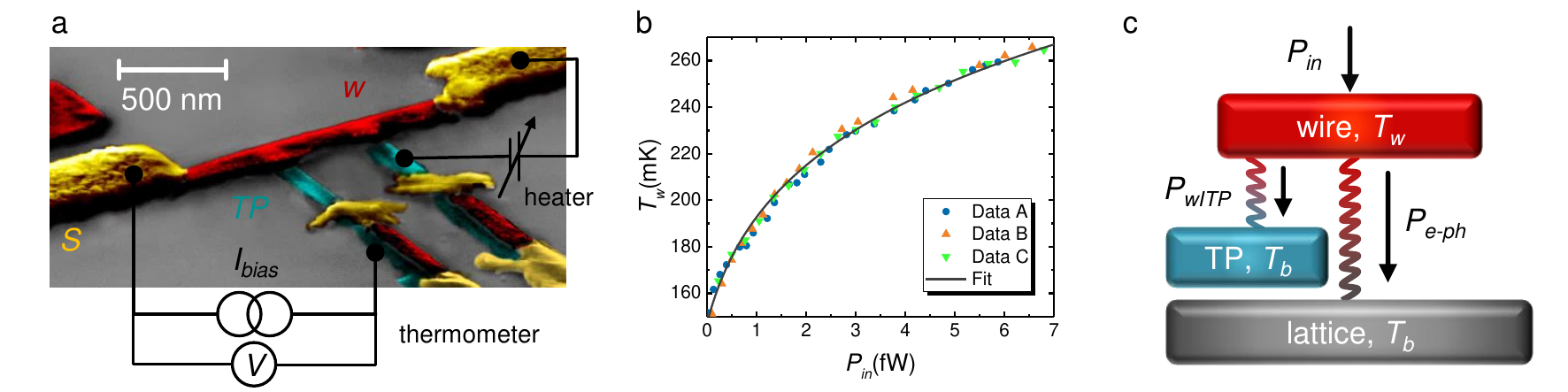}
        \caption{\textbf{Measurement of the electron-phonon coupling constant of the nanowire.} (a) False-color scanning electron micrograph of the AD used to determine $\Sigma_{w}$: Al banks (yellow) and probes (blue) are made of Al and the wire active region (red) is composed of an Al/Cu bilayer. The experimental setup used to obtain $\Sigma_{w}$ is shown as well: a $S$-$w$-$I$-$TP$ JJ is used as heater to inject $P_{in}$ while the other $S$-$w$-$I$-$TP$ JJ acts as thermometer. (b) Electronic temperature of the active region ($T_w$) as a function of the inward power ($P_{in}$): experimental data for three different set of measurements (colored dots and triangles), and fitting curve obtained by solving EQ. \eqref{balance} (gray line). The electron-phonon parameter $\Sigma_w=(1.15\pm 0.02)\times10^{9}$ W/m$^{3}$K$^{5}$ is the resulting fit parameter.
                (c) Thermal model of the active region: the input power ($P_{in}$) is balanced by the outward contributions due to out-diffusion through the thermometer ($P_{wITP}$) and electron-phonon relaxation ($P_{e-ph}$).}
        \label{sigma}
\end{figure*}

\subsection{Measurement of the electron-phonon coupling constant $\Sigma_w$}
In order to obtain the electron-phonon coupling constant of the Al/Cu bilayer ($\Sigma_w$), we injected a power $P_{in}$ directly into the wire and measured its electronic temperature ($T_w$). To this end, we used a voltage-biased $S$-$w$-$I$-$TP$ tunnel junction as heater and a current-biased $S$-$w$-$I$-$TP$ JJ as thermometer \cite{giazotto}, as shown in Fig. \ref{sigma}-a. Since we were interested in the normal-state coupling constant, we performed the experiments at $T_{b}=147.5$ mK ($T_w\ge T_b\ge T_{C,w}$). The resulting $T_w$ versus $P_{in}$ characteristics are shown in Fig. \ref{sigma}-b. As expected, the electronic temperature of the wire monotonically increases by rising the injected power.

In order to analyze the experimental data, we have considered the thermal model shown in Fig. \ref{sigma}-c: the inward contribution of the heater ($P_{in}$) is balanced by the outward power flowing through the thermometer ($P_{wITP}$) and the electron-phonon relaxation channel ($P_{e-ph}$). The outward contributions through the Al banks can be neglected thanks to the Andreev mirror effect \cite{andreevl,andreevll}, since $T_w \ll T_{C,Al}\simeq 1.3$ K.  
As a consequence, the energy balance equation of our system reads
\begin{equation}
P_{in}=P_{wITP}+P_{e-ph}.
\label{balance}
\end{equation}

In a temperature biased JJ, the power flowing from the normal electrode ($w$) to the superconducting tunnel electrode ($TP$) is \cite{giazotto}
\begin{equation}
\begin{split}
P_{wITP}=\\
\frac{1}{e^{2} R_{t}} \int_{-\infty}^{+\infty} \mathrm{d}E E \mathcal{N}_{TP}(E,T_{TP}) \left[ f_{0}(E,T_{w})-f_{0}(E,T_{TP})\right],
\end{split}
\label{NISpower}
\end{equation}
where $\mathcal{N}_{TP}(E, T)=\left| \mathfrak{Re} \left[E+\mathfrak{i}\Gamma/ \sqrt{(E+\mathfrak{i}\Gamma)^2-\Delta_{TP}^2(T)}\right] \right|$ is the smeared normalized Bardeen-Cooper-Schrieffer (BCS) density of state of the superconducting Al electrode and $f_{0}(E,T_{w,TP})=\left[1+\exp{\left(E/k_{B}T_{w,TP}\right)}\right]^{-1}$ is the Fermi-Dirac distribution of the $w$, $TP$ electrode, respectively. $\Gamma=\gamma\Delta_{TP,0}$ is the Dynes parameter accounting for the quasiparticle finite lifetime ($\gamma$ is an empirical parameter, $\Delta_{TP,0}$ is the superconducting energy gap at zero temperature: in our case $\gamma=5\times10^{-3}$ and $\Delta_{TP,0}=$182 $\mu$eV), $\Delta_{TP}(T)$ is the temperature dependent BCS energy gap, $R_{t}$ is the tunnel junction normal-state resistance, $e$ is the electron charge, $k_{B}$ is the Boltzmann constant and $T_{w,TP}$ are the electronic temperatures of the $w$, $TP$ electrode, respectively ($T_{TP}=T_{b}$ due to the big volume of the tunnel probe leads).

The power loss due to the electron-phonon coupling of a normal metal element is \cite{giazotto}
\begin{equation}
P_{e-ph}= \Sigma_w \mathcal{V}_w\left(T_{w}^{5}-T_{b}^{5}\right),
\label{e-phpower}
\end{equation}
where $\mathcal{V}_w$ is the volume of the wire, $T_{w}$ and $T_{b}$ are its electronic and phonon temperature \cite{wellstood}, respectively. 

By solving the Eq. \ref{e-phpower} for $T_{w}$, we computed the expected temperature of the wire active region as a function of $P_{in}$. Since all the other device parameters are known, we fit our experimental data with Eq. \ref{balance} and extracted the values of $\Sigma_w$. The resulting fitting curve is represented by the gray line in Fig. \ref{sigma}-b obtained for $\Sigma_w=(1.15\pm0.02)\times10^{9}$ W/m$^{3}$K$^{5}$. Finally, we notice that, the extracted value of $\Sigma_w$ is in good agreement with the average of the coupling constants of Cu ($\Sigma_{Cu}=2.0\times10^{9}$ W/m$^{3}$K$^{5}$) and Al ($\Sigma_{Al}=0.5\times10^{9}$ W/m$^{3}$K$^{5}$) \cite{giazotto}, weighted on the ratio between the volumes of the two layers ($\Sigma_{w,theo}=1.3\times10^{9}$ W/m$^{3}$K$^{5}$).   

\section{Reproducibility of the Josephson escape sensor}
We investigated the behavior of a second JES. The device is nominally identical to the one presented in the main text. The samples were fabricated simultaneously (same electron beam lithography and angle resolved evaporation steps). Therefore, the differences in the critical current ($I_{C,0-1}\simeq 575$ nA while $I_{C,0-2}\simeq585$ nA) can be ascribed to small dissimilarities in the nanowires width. 

\begin{figure*}[t!]
        \centering
        \includegraphics[width=1\textwidth]{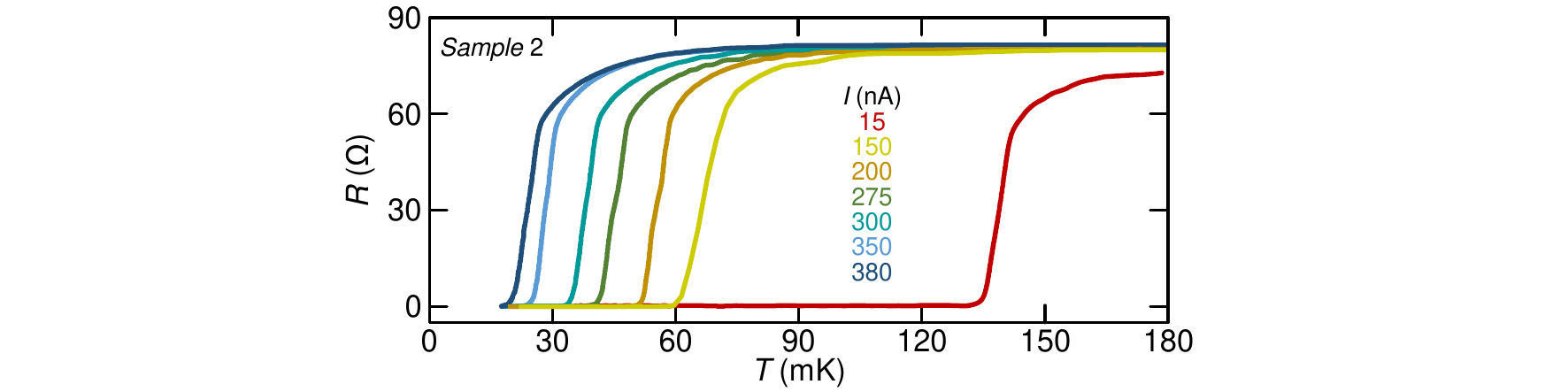}
        \caption{\textbf{Tuning the properties of JES Sample-2.} Selected resistance ($R$) versus temperature $T$ characteristics of another superconducting tunable nanosensor (reported in the main text as Sample-2) measured for different values of bias current ($I$). The $R(T)$ curves exhibit the same dependence on $I$ of Sample-1 (showed in Fig. 3-a of the main text).}
        \label{ResCurrent}
\end{figure*}

The resistance ($R$) versus temperature ($T$) characteristics of the second JES are shown in Fig. \ref{ResCurrent} for different values of bias current ($I$). The device shows a normal state resistance $R_N\simeq 81$ $\Omega$, similar to that of the sample presented in the main text ($\simeq77$ $\Omega$). In agreement with the JES showed in the main text (see Fig. 3-a), by rising $I$ the transition from the superconducting to the dissipative state shifts towards lower temperature. In particular, the transition temperature shows a variation from about 140 mK at $I=15$ nA to about 23 mK at $I=380$ nA. Furthermore, the superconducting-to-dissipative state transition preserves the same appearance for every value of bias current.

The quantitative analysis of the dependence of the effective escape temperature ($T_e^*$) and transition width ($\delta T_C^*$) on the bias current ($I$) is reported in the main text. In particular, the $T_e^*$ and $\delta T_C^*$ dependence on the normalized current ($I/I_C$) of the two JESs are in good agreement, as shown in Fig. 3 of the main text.

\section{Performance of the JES and TES operation}
\begin{figure*}[t!]
        \centering
        \includegraphics[width=1\textwidth]{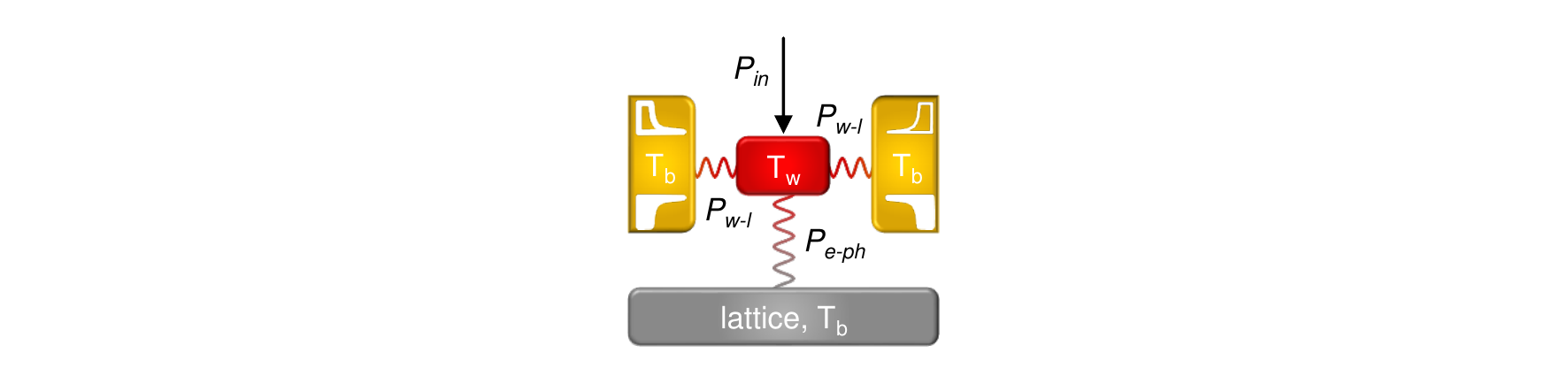}
        \caption{\textbf{Thermal model of the JES.} The input power ($P_{in}$) is balanced by the outward contributions due to out-diffusion through Al leads ($P_{w-l}$) and electron-phonon relaxation ($P_{e-ph}$). Since the electronic temperature of the wire always respects $T_w<0.2T_{C,Al}$, we can consider the two Al leads as perfect Andreev Mirrors and neglect the two $P_{w-l}$ terms.}
        \label{model}
\end{figure*}

The thermal model of the JES is depicted in Fig. \ref{model}. The input power ($P_{in}$) increases the electronic temperature of the one-dimensional JJ ($T_w$). The two superconducting Al leads (kept at the bath temperature $T_b$) serve as Andreev mirrors \cite{andreevl,andreevll}, since $T_w,T_b<0.2\;T_{C,Al}\simeq 1.3$ K. Therefore, the electronic heat out-diffusion ($P_{w,l}$) is exponentially damped by the superconducting energy gap of Al and can be neglected. As a result, the only thermalization channel for the quasiparticles is the electron-phonon coupling ($P_{e-ph}$). 

The JES is operated at $T^*_e(I)$ that can be much smaller than the intrinsic wire critical temperature ($T_C$): the sensor operates strongly in the superconducting state for high values of $I$. Therefore, the electron-phonon coupling is exponentially damped by the presence of the superconducting gap \cite{Timofeev2009,Heikkila}. On the contrary, in TES operation the nanowire is almost in the normal-state ($T\simeq T_C$, so that $\Delta_w=0$), therefore the electron-phonon thermalization can be described by means of the normal metal thin film relation \cite{giazotto}.

In the following we will show all the relations used to extract from our experimental data the principal figures of merit for the JES and the device operated as a conventional TES at about $T_C$. In particular, we will calculate the thermal fluctuations limited noise equivalent power ($NEP_{TFN}$), the frequency resolution ($\delta \nu$) and the thermal time constants (${\tau}_0$ and ${\tau}_{eff}$). 

\subsection{Figures of merit for the TES-mode operation}
For a bolometer the NEP is limited by the thermal fluctuation noise \cite{giazotto}. In case of TES operation ($T\simeq T_C$), the noise equivalent power can be written \cite{Bergmann}\\
\begin{equation}
{NEP}_{TFN,TES} = \sqrt{4\Upsilon {G_{th,TES}} k_{B} {T_C}^{2}}\text{,}
\label{NEP_TES}
\end{equation}
where $\Upsilon = n /(2n+1)$ describes the effect of the temperature gradient across the thermal link (with $n = 5$ for a pure metal), ${G_{th,TES}}$ is the thermal conductance of heat losses, and $k_B$ is the Boltzmann constant.\\
In our device the only channel for thermal losses is the electron phonon coupling ($P_{e-ph}$). Since at $T_C$ the JJ is partially dissipative, the electron-phonon coupling of a normal-metal diffusive thin film ($P_{e-ph, n}$) described by Eq. \ref{e-phpower} can be considered \cite{giazotto}. 

The thermal conductance for a TES (${G_{th}}_{TES}$) is obtained by calculating the derivative of $P_{e-ph, n}$ with respect to the electronic temperature ($T_w$)\cite{Irwin}
\begin{equation}
G_{th,TES} = \dfrac{\text{d} P_{e-ph, n} }{\text{d} T_w} = 5 \Sigma_w {\mathcal{V}}_{w} T_w^4\text{,}
\label{ThermalConductance}
\end{equation}
where $\Sigma_w = 1.15\times 10^{9}$ W m$^{-3}$K$^{-5}$ is the value derived from experimental measures, $\mathcal{V}_{w} = \mathcal{V}_{Al} + \mathcal{V}_{Cu} = 3.83\times 10^{-21}$ m$^{-3}$ is the total volume of the Al/Cu bilayer, with $\mathcal{V}_{Al}$ and $\mathcal{V}_{Cu}$ the aluminum and copper volumes, respectively.\\

In order to determine the performances of a sensor in single-photon detection, the frequency resolution (${\delta\nu}_{TES}$) is the most used figure of merit. For a TES it reads \cite{Irwin}\\
\begin{equation}
{{\delta\nu}_{TES}} = \dfrac{2.36}{\hbar} \sqrt{\dfrac{4}{\alpha} \sqrt{\dfrac{n}{2}} k_B {T_C}^2 {C_{e,TES}}} \text{,}
\label{FrequencyResolution}
\end{equation}
where $\hbar$ is the Planck constant, $\alpha = \dfrac{\text{d} R}{\text{d} T}\dfrac{T}{R}$ is the electrothermal parameter accounting for sharpness of the phase transition from the superconducting to the normal-state \cite{Irwin}, $n=5$ is the electron-phonon coupling for a pure metal and ${C_e}_{TES}$ is the electron heat capacitance. It is interesting to note the strongly dependence on $\alpha$ value which determines the negative electrothermal feedback (NETF) mechanism \cite{Irwin}.\\
The electron heat capacitance is written\\
\begin{equation}
C_{e,TES} = (\gamma_{Cu} \mathcal{V}_{Cu} + \gamma_{Al} \mathcal{V}_{Al}) T_C\text{,}
\label{HeatCapacitance}
\end{equation}
where $\gamma$ is the Sommerfeld coefficient ($\gamma_{Cu}\; =\; 70.5$ JK$^{-2}$m$^{-3}$ , $\gamma_{Al}\; =\; 91$ JK$^{-2}$m$^{-3}$ for copper and aluminum, respectively).
Moreover, for a TES the temperature variation after energy absorption is calculated by solving the time dependent energy balance equation that takes in account all the exchange mechanisms \cite{giazotto}. In particular, the re-thermalization of the quasiparticles to $T_b$ shows an exponential dependence on time with constant
\begin{equation}
\tau = \dfrac{C_{e,TES}}{{G_{th,TES}}}\text{.}
\label{IntrinsicTime}
\end{equation}
This is the intrinsic recovery time of the film which does not consider the Joule heating due to the current flowing through the sensor. Instead, including the heating term in the negative NETF configuration, the pulse recovery time becomes \cite{Irwin}\\
\begin{equation}
\tau_{eff} = \dfrac{\tau}{1 + \dfrac{\alpha}{n}}
\end{equation}
which depends on $\alpha$ (\textit{i.e.} the main parameter of the NETF). When the pulse recovery time is much shorter than the intrinsic time constant ($\tau_{eff}<< \tau$), the energy into the sensor is removed by decreasing its overheating, i.e., compensating for the initial temperature variation (NETF), instead of being dissipated through the substrate.\\
\\

\subsection{Figures of merit for the JES}
The JES operates at the escape temperature $T_e^*$, defined as the maximum temperature measured in the superconducting state before the transition to the dissipative state.
Since the current injection does not change the energy gap of the wire ($\Delta_w\sim const$), the detector works deeply in the superconducting state since $T_e^*(I)\ll T_C$ for high values of $I$. Therefore, all figures of merit have to be calculated deeply in the superconducting state.

The thermal fluctuations limited noise equivalent power can be written
\begin{equation}
{NEP}_{TFN,JES} = \sqrt{4\Upsilon {G_{th,JES}} k_{B} {T_e^*}^{2}}\text{.}
\label{NEP_JE}
\end{equation}

The thermal conductance in the superconducting state is described by \cite{Heikkila}
\begin{equation}
\begin{split}
{G_{th,JES}} \approx \\
\dfrac{\Sigma_w \mathcal{V}_{m} {T_e^*}^4}{96 \varsigma(5)}\left[ f_1 \left(\dfrac{1}{\tilde{\Delta}}\right) \cosh(\tilde{h}) e^{-\tilde{\Delta}} + \pi \tilde{\Delta}^5 f_2 (\dfrac{1}{\tilde\Delta})e^{-2 \tilde{\Delta}}  \right] \text{,}
\end{split}
\end{equation}
where the first and the second terms refer to electron-phonon scattering and recombination processes, respectively. Here, $\varsigma(5)= 1.0369$ is the Riemann zeta function, $\tilde{\Delta} = \Delta_w/k_B T$ (with $\Delta_w = 23$ $\mu$eV the experimental gap of the wire), $\tilde{h} = h/k_B T$ is exchange field (0 in this case), $f_1(x) = \begin{matrix} \sum_{n=0}^3  C_n x^n \end{matrix} $ with $C_0 \approx 440, C_1 \approx 500, C_2 \approx 1400, C_3 \approx 4700$ and $f_2(x) = \begin{matrix} \sum_{n=0}^2  B_n x^n \end{matrix}$ with $B_0 = 64, B_1 = 144, B_2= 258$.\\
Operating as a calorimeter, the frequency resolution $\delta\nu_{JE}$ of a superconducting thermal sensor can be computed from \cite{Virtanen}\\
\begin{equation}
\delta\nu_{JES} = \dfrac{4}{\hbar}\sqrt{2 \ln{2} k_B {T_e^*}^2 {{C_{e,JES}}}}\text{,}
\label{FreqResSES}
\end{equation}
where ${C_{e,JES}}$ is the electron heat capacitance calculated at the escape temperature $T_e^*$ considering the damping term $\Theta_{Damp}$ typical in a BCS superconductor\\
\begin{equation}
{C_{e,JES}} = (\gamma_{Cu} \mathcal{V}_{Cu} + \gamma_{Al} \mathcal{V}_{Al})T_e^* \Theta_{Damp}\text{.}
\end{equation}
The low temperature exponential suppression $\Gamma_{Damp}$ with respect to the normal state heat capacitance, which leads to the JES high detection sensitivity, is written as \cite{Rabani}\\
\begin{equation}
\Theta_{Damp} = \dfrac{C_s}{1.34(\gamma_{Cu} + \gamma_{Al})T_e^*}\text{.}
\end{equation}
The electronic heat capacitance is given by
\begin{equation}
C_s = 1.34 (\gamma_{Cu} + \gamma_{Al})T_C \left( \dfrac{\Delta_w}{k_B T_e^*} \right)^{3/2} e^{\Delta_w/k_B T_e^*}\text{,}
\end{equation}
where the critical temperature $T_{C,w}\simeq 150$ mK is related to the measured gap $\Delta_w\simeq23$ $\mu$eV.\\
Considering the relaxation time in the weak link, the predominant thermalization mechanism is due to the electron-phonon interaction, which defines the sensor relaxation half-time $\tau_{1/2}$ \cite{Virtanen}\\
\begin{equation}
\tau_{1/2} = \tau \ln{2} \text{,}
\label{TauMezzi} 
\end{equation}
where $\tau$ is the thermal time constant (see Eq. \ref{IntrinsicTime}) considering ${C_{e,JES}}$ and ${G_{th,JES}}$ for a Josephson escape sensor.

\end{document}